\documentclass[aps,groupedaddress,showpacs,floatfix]{revtex4}
\usepackage{amssymb}
\usepackage{amsmath}
\usepackage{graphicx}
\DeclareMathOperator{\sgn}{sgn}
\DeclareMathOperator{\Ai}{Ai}
\begin{document}

\title{Replica Bethe ansatz derivation of the GOE Tracy-Widom distribution
in one-dimensional directed polymers with free boundary conditions}

\author{Victor Dotsenko }

\affiliation{LPTMC, Universit\'e Paris VI, 75252 Paris, France}

\affiliation{L.D.\ Landau Institute for Theoretical Physics,
 119334 Moscow, Russia}

\date{\today}

\begin{abstract}

 The distribution function of the free energy fluctuations
in one-dimensional directed polymers with free boundary conditions
is derived by mapping the
replicated problem to the $N$-particle quantum boson system with attractive
interactions.   It is shown that in the thermodynamic limit this function
is described by the universal Tracy-Widom distribution
of the Gaussian orthogonal ensemble.

\end{abstract}


\maketitle

\medskip

\section{Introduction}

Directed polymers in a quenched random potential have
been the subject of intense investigations during the past two
decades (see e.g. \cite{hh_zhang_95,burgers_74,hhf_85,numer1,numer2,kardar_87}).
In the  one-dimensional case
we deal with an elastic string  directed along the $\tau$-axis
within an interval $[0,t]$. Randomness enters the problem
through a disorder potential $V[\phi(\tau),\tau]$, which competes against
the elastic energy.  The problem is defined by the Hamiltonian
\begin{equation}
   \label{1}
   H[\phi(\tau), V] = \int_{0}^{t} d\tau
   \Bigl\{\frac{1}{2} \bigl[\partial_\tau \phi(\tau)\bigr]^2
   + V[\phi(\tau),\tau]\Bigr\};
\end{equation}
where the disorder potential $V[\phi,\tau]$
is Gaussian distributed with a zero mean $\overline{V(\phi,\tau)}=0$
and the $\delta$-correlations:
\begin{equation}
   \label{2}
{\overline{V(\phi,\tau)V(\phi',\tau')}} = u \delta(\tau-\tau') \delta(\phi-\phi')
\end{equation}
Here the parameter $u$ describes the strength of the disorder.
Note that such system is equivalent to the problem of the
Kardar-Parisi-Zang (KPZ) equation describing the growth in time of an interface
in the presence of noise \cite{KPZ}.

In what follows we consider the problem in which the polymer is fixed
at the origin, $\phi(0)=0$ and it is free at $\tau = t$.
In other words, for a given realization of the random potential
$V$ the partition function of the considered system is:
\begin{equation}
\label{3}
   Z = \int_{-\infty}^{+\infty} dx \; Z(x) \; = \; \exp\{-\beta F\}
\end{equation}
where
\begin{equation}
\label{4}
   Z(x) = \int_{\phi(0)=0}^{\phi(t)=x}
              {\cal D} \phi(\tau)  \;  \mbox{\Large e}^{-\beta H[\phi]}
\end{equation}
is the partition function of the system with the fixed boundary conditions,
$\phi(0)=0$ and $\phi(t)=x$ and
where $F$ is the total free energy.
Besides the usual extensive   part $f_{0} t$
(where $f_{0}$ is the linear free energy density),
the total free energy $F$ of such system is known to contain the disorder dependent
fluctuating contribution $\tilde{F}$. In the limit of large $t$
the typical value of the free energy fluctuations scales with $t$ as
$\tilde{F}  \propto t^{1/3}$ (see e.g. \cite{hhf_85,numer1,numer2,kardar_87}).
In other words,  the total  free energy of the system
can be represented as
\begin{equation}
\label{5}
F \; = \; f_{0} t \; + \; c \, t^{1/3}\; f
\end{equation}
where $c$ is a non-universal parameter, which depends on the temperature and the strength of disorder, and $f$ is the random quantity which in the thermodynamic
limit $t\to\infty$ is described by a non-trivial universal
distribution function $P(f)$. Note that according to eqs.(\ref{3})-(\ref{5}),
the trivial self-averaging  contribution $f_{0}t$ to the free energy can be eliminated
by a simple redefinition of the partition function:
\begin{equation}
\label{6}
Z \; =\; \exp\{-\beta f_{0} t\} \, \tilde{Z}
\end{equation}
so that
\begin{equation}
\label{7}
\tilde{Z} \; = \; \exp\{-\lambda f\}
\end{equation}
where
\begin{equation}
\label{8}
\lambda \; = \; \beta \, c \, t^{1/3}
\end{equation}

For the similar problem with the zero boundary conditions, $\phi(0)=\phi(t)=0$,
the corresponding distribution function was proved to be described by
the Gaussian Unitary Ensemble (GUE) Tracy-Widom distribution
\cite{KPZ-TW1,KPZ-TW2,BA-replicas,LeDoussal1}. In the course  of this proof
rather efficient Bethe ansatz replica technique has been developed
\cite{BA-replicas,LeDoussal1}. In particular, in terms of this technique
the corresponding multi-point free energy distribution functions
have been derived \cite{Prolhac-Spohn}.
Recently,  the free energy
distribution function for the directed polymer problem with
the free boundary conditions, eqs.(\ref{1})-(\ref{4}), has been obtained
\cite{LeDoussal2}. It was shown that the function $P(f)$ is  the
Gaussian Orthogonal Ensemble (GOE) Tracy-Widom distribution.
In this paper I would like to present sufficiently simple alternative way
of derivation of the same result which does not require rather
complicated technique of the Fredholm Pfaffian described in \cite{LeDoussal2}.

Let us introduce the function
\begin{equation}
\label{9}
W(f) \; \equiv \; \int_{f}^{\infty} \; df' \; P(f')
\end{equation}
which gives the probability that the random free energy
is bigger that a given value $f$.
It will be shown that in the thermodynamic limit, $t \to \infty$,
this function is equal to the Fredholm
determinant
\begin{equation}
\label{10}
W(f) \; = \; \det(1 - \hat{K}_{-f}) \; \equiv \; F_{1}(-f)
\end{equation}
with the kernel
\begin{equation}
 \label{11}
K_{-f}(\omega, \omega') \; = \; \Ai(\omega + \omega' - f) \; ; \; \; \;  \; \; \;  \; \; \;
(\omega, \omega' \; > 0)
\end{equation}
which is the GOE Tracy-Widom distribution \cite{TW2,Ferrari-Spohn}.
Explicitly,
\begin{equation}
 \label{11a}
F_{1}(s) \; = \;
\exp\Biggl[-\frac{1}{2}\int_{s}^{+\infty} d\xi \; (\xi-s) \; q^{2}(\xi) \;
-\frac{1}{2}\int_{s}^{+\infty} d\xi \; q(\xi)\Biggr]
\end{equation}
where $q(\xi)$ is the solution of the Panlev\'e II differential equation,
$q''(\xi)=\xi q(\xi)+2 q^{3}(\xi)$,
with the boundary condition  $q(\xi\to+\infty)=\Ai(\xi)$.

It should be noted that the present paper is rather technical. The main message of this
work is not the final result itself (which is well known anyway) but the presentation
of the general method and  new technical tricks used in the derivation.
Section II is devoted to the standard reformulation of the considered problem in terms of
one-dimensional $N$-particle system of quantum bosons with attractive
$\delta$-interactions \cite{kardar_87} .
Here it is shown that the calculation of the the free energy probability
distribution function, eq.(\ref{9}), reduces to the
summation over all the spectrum of eigenstates of this $N$-particle
problem. This summation is performed in Section III, where in the thermodynamic
limit, $t\to\infty$, the result, eqs.(\ref{10})-(\ref{11}) is derived.
The concluding remarks and as well as the key points of the calculations are listed
in the final Section IV.


\section{Mapping to quantum bosons}

In terms of the partition function $\tilde{Z}$, eq.(\ref{7}),
the function $W(f)$, eq.(\ref{9}), can be defined as follows:
\begin{equation}
 \label{12}
W(f) = \lim_{\lambda\to\infty} \sum_{N=0}^{\infty} \frac{(-1)^{N}}{N!}
\exp(\lambda N f) \; \overline{\tilde{Z}^{N}}
\end{equation}
where $\overline{(...)}$ denotes the averaging over quenched disorder.
Indeed, substituting here eq.(\ref{7}), we have
\begin{eqnarray}
 \label{13}
W(f) &=& \lim_{\lambda\to\infty} \sum_{N=0}^{\infty} \frac{(-1)^{N}}{N!}
\int_{-\infty}^{+\infty} \; df' \; P(f')
\exp\{\lambda N (f - f') \}
\\
\nonumber
\\
\nonumber
&=& \lim_{\lambda\to\infty}
\int_{-\infty}^{+\infty} \; df' \; P(f')
\exp\bigl[-\exp\{\lambda (f - f') \} \bigr]
\\
\nonumber
\\
\nonumber
&=&
\int_{-\infty}^{+\infty} \; df' \; P(f') \;
\theta\bigl(f-f'\bigr)
\end{eqnarray}
which coincides with the definition, eq.(\ref{9}).

Later on we will see that the integration over $x$ in the definition of the
partition function, eq.(\ref{3}), requires proper regularization
at both limits $\pm\infty$. For that reason it is convenient to represent it
in the form of two contributions:
\begin{equation}
\label{14}
Z \; = \;
\int_{-\infty}^{0} dx \; Z(x) \; + \;
\int_{0}^{+\infty} dx \; Z(x) \; \equiv \;
Z_{(-)} \; + \; Z_{(+)}
\end{equation}
Thus, taking into account the definition
eq.(\ref{6}), we get
\begin{eqnarray}
 \nonumber
W(f) &=& \lim_{\lambda\to\infty} \sum_{N=0}^{\infty} \frac{(-1)^{N}}{N!}
\exp\{\lambda N f +\beta N f_{0} t\}
\overline{\bigl( Z_{(-)} \; + \; Z_{(+)} \bigr)^{N}}
\\
\nonumber
\\
\nonumber
&=&
\lim_{\lambda\to\infty} \sum_{K,L=0}^{\infty} \frac{(-1)^{K+L}}{K! \, L!}
\exp\{\lambda (K+L) f +\beta (K+L) f_{0} t\} \;
\overline{Z_{(-)}^{K}  Z_{(+)}^{L} }
\\
\nonumber
\\
\nonumber
&=&
\lim_{\lambda\to\infty} \sum_{K,L=0}^{\infty} \frac{(-1)^{K+L}}{K! \, L!}
\exp\{\lambda (K+L) f +\beta (K+L) f_{0} t\} \times
\\
\nonumber
\\
&\times&
\int_{-\infty}^{0} dx_{1}...dx_{K}
\int_{0}^{+\infty} dy_{1}...dy_{L}
\Psi(x_{1},...,x_{K},y_{L},...,y_{1} ; t)
\label{15}
\end{eqnarray}
where
\begin{equation}
\label{16}
\Psi(x_{1}, ..., x_{N} ; t) \; \equiv \;
\overline{Z(x_{1}) \, Z(x_{2}) \, ... \, Z(x_{N})}
\end{equation}
Using the relations, eqs.(\ref{1}), (\ref{2}) and (\ref{4}), after simple Gaussian
averaging we obtain
\begin{equation}
   \label{17}
    \Psi(x_{1}, ..., x_{N} ; t) \; = \;
\prod_{a=1}^{N} \Biggl[\int_{\phi_a(0)=0}^{\phi_a(t)=x_a} {\cal D} \phi_a(\tau)\Biggr]
  \;  \exp\bigl(-\beta  H_{N} [\phi_{1}, \phi_{2}, ..., \phi_{N}] \bigr)
\end{equation}
where
\begin{equation}
\label{18}
   H_{N} [\phi_{1}, \phi_{2}, ..., \phi_{N}] \; = \;
   \frac{1}{2} \int_{0}^{t} d\tau \Biggl(
   \sum_{a=1}^{N} \bigl[\partial_\tau\phi_{a}(\tau)\bigr]^2
   - \beta u \sum_{a\not= b}^{N}
   \delta\bigl[\phi_{a}(\tau)-\phi_{b}(\tau)\bigr] \Biggr)
\end{equation}
The propagator $\Psi({\bf x}; t)$, eq.(\ref{17}), describes $N$ trajectories 
$\phi_{a}(\tau)$ all starting at zero ($\phi_{a}(0) = 0$), and coming to $N$ different points $\{x_{1}, ..., x_{N}\}$
at $\tau = t$. One can easily show that $\Psi({\bf x}; t)$
can be obtained as the solution of the  linear differential equation
\begin{equation}
   \label{19}
\beta \, \partial_t \Psi({\bf x}; t) \; = \;
\frac{1}{2}\sum_{a=1}^{N}\partial_{x_a}^2 \Psi({\bf x}; t)
  \; + \; \frac{1}{2} \, \kappa \sum_{a\not=b}^{N} \delta(x_a-x_b) \Psi({\bf x}; t)
\end{equation}
with the initial condition
\begin{equation}
   \label{20}
\Psi({\bf x}; 0) = \Pi_{a=1}^{N} \delta(x_a)
\end{equation}
and the interaction parameter $\kappa = \beta^{3}u$. 
One can easily see that Eq.(\ref{19}) is the imaginary-time
Schr\"odinger equation
\begin{equation}
   \label{21}
-\beta \, \partial_t \Psi({\bf x}; t) = \hat{H} \Psi({\bf x}; t)
\end{equation}
with the Hamiltonian
\begin{equation}
   \label{22}
   \hat{H} =
    -\frac{1}{2}\sum_{a=1}^{N}\partial_{x_a}^2
   -\frac{1}{2}\, \kappa \sum_{a\not=b}^{N} \delta(x_a-x_b)
\end{equation}
which describes  $N$ bose-particles  interacting via
the {\it attractive} two-body potential $-\kappa \delta(x)$.
A generic  eigenstate of such system is characterized by $N$ momenta
$\{ q_{a} \} \; (a=1,...,N)$ which are splitted into
$M$  ($1 \leq M \leq N$) "clusters" described by
continuous real momenta $q_{\alpha}$ $(\alpha = 1,...,M)$
and having $n_{\alpha}$ discrete imaginary "components"
(for details see \cite{Lieb-Liniger,McGuire,Yang,Calabrese,BA-replicas,rev-TW}):
\begin{equation}
   \label{23}
q_{a} \; \equiv \; q^{\alpha}_{r} \; = \;
q_{\alpha} - \frac{i\kappa}{2}  (n_{\alpha} + 1 - 2r)
\;\; ; \; \;\;\; \;\;\; \;\;\;
(r = 1, ..., n_{\alpha})
\end{equation}
with the constraint
\begin{equation}
   \label{24}
\sum_{\alpha=1}^{M} n_{\alpha} = N
\end{equation}
A generic solution  $\Psi({\bf x},t)$
of the Schr\"odinger equation (\ref{19}) with the initial conditions, Eq.(\ref{20}),
can be represented in the form of the linear combination of the eigenfunctions
$\Psi_{\bf q}^{(M)}({\bf x})$:
\begin{equation}
   \label{26}
\Psi(x_{1}, ..., x_{N}; t) =
\sum_{M=1}^{N} \frac{1}{M!} \Biggl[\int {\cal D}^{(M)} ({\bf q},{\bf n})\Biggr] \;
|C_{M}({\bf q},{\bf n})|^{2} \;
\Psi^{(M)}_{{\bf q}}({\bf x})
{\Psi^{(M)}_{{\bf q}}}^{*}({\bf 0}) \;
\exp\bigl\{-E_{M}({\bf q}) t \bigr\}
\end{equation}
where we have introduced the notation
\begin{equation}
   \label{27}
\int {\cal D}^{(M)} ({\bf q},{\bf n}) \equiv
\prod_{\alpha=1}^{M} \Biggl[\int_{-\infty}^{+\infty} \frac{dq_{\alpha}}{2\pi} \sum_{n_{\alpha}=1}^{\infty}\Biggr]
{\boldsymbol \delta}\Bigl(\sum_{\alpha=1}^{M} n_{\alpha} \; , \;  N\Bigr)
\end{equation}
and ${\boldsymbol \delta}(k , m)$ is the Kronecker symbol; 
note that the presence of this Kronecker symbol in the above equation 
allows to extend the summations over $n_{\alpha}$'s to infinity.
Here (non-normalized) eigenfunctions are \cite{BA-replicas,rev-TW}
\begin{equation}
\label{28}
\Psi^{(M)}_{{\bf q}}({\bf x}) =
\sum_{{\cal P}}  \;
\prod_{a<b}^{N}
\Biggl[
1 +i \kappa \frac{\sgn(x_{a}-x_{b})}{q_{{\cal P}_a} - q_{{\cal P}_b}}
\Biggr] \;
\exp\Bigl[i \sum_{a=1}^{N} q_{{\cal P}_{a}} x_{a} \Bigr]
\end{equation}
where the summation goes over $N!$ permutations ${\cal P}$ of $N$ momenta $q_{a}$,
eq.(\ref{23}),  over $N$ particles $x_{a}$;
the normalization factor
\begin{equation}
   \label{29}
|C_{M}({\bf q}, {\bf n})|^{2} = \frac{\kappa^{N}}{N! \prod_{\alpha=1}^{M}\bigl(\kappa n_{\alpha}\bigr)}
\prod_{\alpha<\beta}^{M}
\frac{\big|q_{\alpha}-q_{\beta} -\frac{i\kappa}{2}(n_{\alpha}-n_{\beta})\big|^{2}}{
      \big|q_{\alpha}-q_{\beta} -\frac{i\kappa}{2}(n_{\alpha}+n_{\beta})\big|^{2}}
\end{equation}
and the eigenvalues:
\begin{equation}
\label{30}
E_{M}({\bf q}) \; = \;
\frac{1}{2\beta} \sum_{\alpha=1}^{N} q_{a}^{2} \; = \;
 \frac{1}{2\beta} \sum_{\alpha=1}^{M} \; n_{\alpha} q_{\alpha}^{2}
- \frac{\kappa^{2}}{24\beta}\sum_{\alpha=1}^{M} (n_{\alpha}^{3}-n_{\alpha})
\end{equation}
Note that the eigenfunctions, eq.(\ref{28}), are symmetric with respect
to permutations of  all its arguments $x_{1}, ..., x_{N}$, and
\begin{equation}
\label{31}
\Psi^{(M)}_{{\bf q}}({\bf 0}) = N!
\end{equation}
In this way the problem of the calculation of the free energy probability
distribution function, eq.(\ref{15}), reduces to the 
summation over all the spectrum of the eigenstates of the $N$-particle
bosonic problem, which is parametrized by the set of both the continuous,
$\{q_{1}, ..., q_{M}\}$, and the discrete
$\{n_{1}, ...,n_{M}\}; \; (M = 1, ..., N); \; (N = 1, ..., \infty)$
degrees of freedom.

\section{Free energy probability distribution function}

Substituting eqs.(\ref{26})-(\ref{31}) into eq.(\ref{15}),
(defining $f_{0} = \frac{1}{24}\beta^4 u^2 $, the factor $f_{0}$ drops out of the further calculations) 
we get:
\begin{eqnarray}
 \nonumber
W(f) &=& 1 + \lim_{\lambda\to\infty}
\sum_{K+L\geq 1}^{\infty} \;(-1)^{K+L} \;
\mbox{\LARGE e}^{\lambda (K+L) f}
\times
\\
\nonumber
\\
&\times&
\sum_{M=1}^{K+L} \frac{1}{M!}
\prod_{\alpha=1}^{M}
\Biggl[
\sum_{n_{\alpha}=1}^{\infty}
\int_{-\infty}^{+\infty} \frac{dq_{\alpha}}{2\pi \kappa n_{\alpha}} \kappa^{n_{\alpha}}
\mbox{\LARGE e}^{-\frac{t}{2\beta} n_{\alpha} q_{\alpha}^{2}
+ \frac{\kappa^{2}}{24 \beta} n_{\alpha}^{3}}
\Biggr]
\; {\boldsymbol \delta}\Bigl(\sum_{\alpha=1}^{M} n_{\alpha} \; , \;  N\Bigr)
\; |\tilde{C}_{M}({\bf q}, {\bf n})|^{2}
\; I_{K,L} ({\bf q}, {\bf n})
\label{32}
\end{eqnarray}
where
\begin{equation}
   \label{29a}
|\tilde{C}_{M}({\bf q}, {\bf n})|^{2} \; = \;
\prod_{\alpha<\beta}^{M}
\frac{\big|q_{\alpha}-q_{\beta} -\frac{i\kappa}{2}(n_{\alpha}-n_{\beta})\big|^{2}}{
      \big|q_{\alpha}-q_{\beta} -\frac{i\kappa}{2}(n_{\alpha}+n_{\beta})\big|^{2}}
\end{equation}
and
\begin{eqnarray}
\nonumber
I_{K,L} ({\bf q}, {\bf n}) &=&
\sum_{{\cal P}^{(K,L)}}  \sum_{{\cal P}^{(K)}} \sum_{{\cal P}^{(L)}} \;
\prod_{a=1}^{K} \prod_{c=1}^{L}
\Biggl[
\frac{q_{{\cal P}_a^{(K)}} - q_{{\cal P}_c^{(L)}} - i \kappa}{q_{{\cal P}_a^{(K)}} - q_{{\cal P}_c^{(L)}}}
\Biggr]
\times
\prod_{a<b}^{K}\Biggl[\frac{q_{{\cal P}_a^{(K)}} - q_{{\cal P}_b^{(K)}}  - i \kappa }{q_{{\cal P}_a^{(K)}} - q_{{\cal P}_b^{(K)}}}\Biggr]
\times
\prod_{c<d}^{L}\Biggl[\frac{q_{{\cal P}_c^{(L)}} - q_{{\cal P}_d^{(L)}}  + i \kappa }{q_{{\cal P}_c^{(L)}} - q_{{\cal P}_d^{(L)}}}\Biggr]
\times
\\
\nonumber
\\
\nonumber
&\times&
\int_{-\infty < x_{1} \leq ... \leq x_{K}\leq 0} dx_{1} ... dx_{K} \;
\exp\Bigl[i \sum_{a=1}^{K} (q_{{\cal P}_{a}^{(K)}}-i\epsilon) x_{a} \Bigr]
\\
\nonumber
\\
&\times&
\int_{0 \leq y_{L} \leq ... \leq y_{1} < +\infty} dy_{L} ... dy_{1} \;
\exp\Bigl[i \sum_{c=1}^{L} (q_{{\cal P}_{c}^{(L)}}+i\epsilon) y_{c} \Bigr]
\label{33}
\end{eqnarray}
Here the summation over all permutations ${\cal P}$ of $(K+L)$ momenta
$\{q_{1}, ..., q_{K+L}\}$  over $K$ "negative" particles
$\{x_{1}, ..., x_{K}\}$
and $L$ "positive" particles $\{y_{L}, ..., y_{1}\}$
are divided   into three parts: the permutations ${\cal P}^{(K)}$
of $K$ momenta (taken at random out of the total list $\{q_{1}, ..., q_{K+L}\}$)
over $K$ "negative" particles, the permutations ${\cal P}^{(L)}$
of the remaining $L$ momenta over $L$ "positive" particles, and
finally the permutations ${\cal P}^{(K,L)}$ (or the exchange) of the
momenta between the group $"K"$ and the group $"L"$.
Note also that the integrations both over $x_{a}$'s and over $y_{c}$'s
in eq.(\ref{33}) require proper regularization at $-\infty$ and $+\infty$ correspondingly.
This is done in the standard way by introducing a supplementary parameter $\epsilon$
which will be set to zero in final results. The result of the
integrations can be represented as follows:
\begin{eqnarray}
\nonumber
I_{K,L} ({\bf q}, {\bf n}) &=& i^{-(K+L)}
\sum_{{\cal P}^{(K,L)}}  \; \;
\prod_{a=1}^{K} \prod_{c=1}^{L}
\Biggl[
\frac{q_{{\cal P}_a^{(K)}} - q_{{\cal P}_c^{(L)}} - i \kappa}{q_{{\cal P}_a^{(K)}} - q_{{\cal P}_c^{(L)}}}
\Biggr]
\times
\\
\nonumber
\\
\nonumber
&\times&
\sum_{{\cal P}^{(K)}} \; \;
\frac{1}{q^{(-)}_{{\cal P}_{1}^{(K)}} \bigl(q^{(-)}_{{\cal P}_{1}^{(K)}} + q^{(-)}_{{\cal P}_{2}^{(K)}}\bigr)... \bigl(q^{(-)}_{{\cal P}_{1}^{(K)}} + ... + q^{(-)}_{{\cal P}_{K}^{(K)}}\bigr)}
\prod_{a<b}^{K}\Biggl[\frac{q^{(-)}_{{\cal P}_a^{(K)}} - q^{(-)}_{{\cal P}_b^{(K)}}  - i \kappa }{q^{(-)}_{{\cal P}_a^{(K)}} - q^{(-)}_{{\cal P}_b^{(K)}}}\Biggr]
\times
\\
\nonumber
\\
&\times&
\sum_{{\cal P}^{(L)}} \; \;
\frac{(-1)^{L}}{q^{(+)}_{{\cal P}_{1}^{(L)}} \bigl(q^{(+)}_{{\cal P}_{1}^{(L)}} + q^{(+)}_{{\cal P}_{2}^{(L)}}\bigr)... \bigl(q^{(+)}_{{\cal P}_{1}^{(L)}} + ... + q^{(+)}_{{\cal P}_{L}^{(L)}}\bigr)}
\prod_{c<d}^{L}\Biggl[\frac{q^{(+)}_{{\cal P}_c^{(L)}} - q^{(+)}_{{\cal P}_d^{(L)}}  + i \kappa }{q^{(+)}_{{\cal P}_c^{(L)}} - q^{(+)}_{{\cal P}_d^{(L)}}}\Biggr]
\label{34}
\end{eqnarray}
where
\begin{equation}
\label{34a}
q^{(\pm)}_a \; \equiv \; q_{a} \pm i\epsilon
\end{equation}
Using the "magic" Bethe ansatz combinatorial identity \cite{LeDoussal2},
\begin{equation}
 \label{35}
\sum_{P} \frac{1}{q_{p_{1}} (q_{p_{1}} + q_{p_{2}})... (q_{p_{1}} + ... + q _{p_{N}})}
\prod_{a<b}^{N}\Biggl[\frac{q_{p_a} - q_{p_b}  - i \kappa }{q_{p_a} - q_{p_b}}\Biggr] \; = \;
\frac{1}{\prod_{a=1}^{N} q_{a}} \;
\prod_{a<b}^{N}\Biggl[\frac{q_{a} + q_{b}  + i \kappa }{q_{a} + q_{b}}\Biggr]
\end{equation}
(where the summation goes over all permutations $P$ of $N$ momenta $\{q_{1}, ..., q_{N}\}$) we get:
\begin{eqnarray}
\nonumber
I_{K,L} ({\bf q}, {\bf n}) &=& i^{-(K+L)}
\sum_{{\cal P}^{(K,L)}}  \; \;
\prod_{a=1}^{K} \prod_{c=1}^{L}
\Biggl[
\frac{q_{{\cal P}_a^{(K)}} - q_{{\cal P}_c^{(L)}} - i \kappa}{q_{{\cal P}_a^{(K)}} - q_{{\cal P}_c^{(L)}}}
\Biggr]
\times
\\
\nonumber
\\
&\times&
\frac{1}{\prod_{a=1}^{K} q^{(-)}_{{\cal P}_{a}^{(K)}} }
\prod_{a<b}^{K}\Biggl[\frac{q^{(-)}_{{\cal P}_a^{(K)}} + q^{(-)}_{{\cal P}_b^{(K)}}  + i \kappa }{q^{(-)}_{{\cal P}_a^{(K)}} + q^{(-)}_{{\cal P}_b^{(K)}}}\Biggr]
\times
\frac{(-1)^{L}}{\prod_{c=1}^{L} q^{(+)}_{{\cal P}_{c}^{(L)}} }
\prod_{c<d}^{L}\Biggl[\frac{q^{(+)}_{{\cal P}_c^{(L)}} + q^{(+)}_{{\cal P}_d^{(L)}}  - i \kappa }{q^{(+)}_{{\cal P}_c^{(L)}} + q^{(+)}_{{\cal P}_d^{(L)}}}\Biggr]
\label{36}
\end{eqnarray}
Further simplification comes from one important property of the
Bethe ansatz wave function, eq.(\ref{28}). It has such structure that
for ordered particles positions (e.g. $x_{1}<x_{2}<...<x_{N}$)
in the summation over permutations the momenta $q_{a}$ belonging
to the same cluster also remain ordered. In other words,
if we consider the momenta, eq.(\ref{23}), of a cluster $\alpha$, 
$\{q_{1}^{\alpha}, q_{2}^{\alpha}, ..., q_{n_{\alpha}}^{\alpha}\}$,
belonging correspondingly to the particles $\{x_{i_{1}} < x_{i_{2}} < ... < x_{i_{n_{\alpha}}}\}$, 
the permutation of any two momenta $q_{r}^{\alpha}$
and $q_{r'}^{\alpha}$ of this {\it ordered} set gives zero contribution.
Thus, in order to perform the summation over the permutations ${\cal P}^{(K,L)}$
in eq.(\ref{36}) it is sufficient to split the momenta of each cluster into two parts:
$\{q_{1}^{\alpha}, ...,  q_{m_{\alpha}}^{\alpha} ||
q_{m_{\alpha}+1}^{\alpha}..., q_{n_{\alpha}}^{\alpha}\}$, where $m_{\alpha} = 0, 1, ..., n_{\alpha}$ and 
where the momenta $q_{1}^{\alpha}, ...,  q_{m_{\alpha}}^{\alpha}$ belong to the particles
of the sector $"K"$, while the momenta $q_{m_{\alpha}+1}^{\alpha}..., q_{n_{\alpha}}^{\alpha}$ 
belong to the particles of the sector $"L"$. 

Let us introduce the numbering of the momenta
of the sector $"L"$ in the reversed order:
\begin{eqnarray}
\nonumber
q_{n_{\alpha}}^{\alpha} &\to&  {q^{*}}_{1}^{\alpha}
\\
\nonumber
q_{n_{\alpha}-1}^{\alpha} &\to&  {q^{*}}_{2}^{\alpha}
\\
\nonumber
&........&
\\
q_{m_{\alpha}+1}^{\alpha} &\to&  {q^{*}}_{s_{\alpha}}^{\alpha}
\label{37}
\end{eqnarray}
where $m_{\alpha} + s_{\alpha} = n_{\alpha}$ and (s.f. eq.(\ref{23}))
\begin{equation}
\label{38}
{q^{*}}_{r}^{\alpha} \; = \; q_{\alpha} + \frac{i \kappa}{2} (n_{\alpha} + 1 - 2r)
\; = \; q_{\alpha} + \frac{i \kappa}{2} (m_{\alpha} + s_{\alpha} + 1 - 2r)
\end{equation}
By definition, the integer parameters $\{m_{\alpha}\}$ and $\{s_{\alpha}\}$
fulfill the global constrains
\begin{eqnarray}
\label{39}
\sum_{\alpha=1}^{M} m_{\alpha} &=& K
\\
\nonumber
\\
\sum_{\alpha=1}^{M} s_{\alpha} &=& L
\label{40}
\end{eqnarray}
In this way the summation over permutations ${\cal P}^{(K,L)}$
in eq.(\ref{36}) is changed by the summations over the integer parameters
$\{m_{\alpha}\}$ and $\{s_{\alpha}\}$:
\begin{equation}
\label{41}
\sum_{{\cal P}^{(K,L)}} \; \bigl( ... \bigr) \; \to \;
\prod_{\alpha=1}^{M}
\Biggl[
\sum_{m_{\alpha}+s_{\alpha} \geq 1}^{\infty} \;
{\boldsymbol \delta}\Bigl(m_{\alpha}+s_{\alpha} \; , \;  n_{\alpha}\Bigr)
\Biggr] \;
{\boldsymbol \delta}\Bigl(\sum_{\alpha=1}^{M} m_{\alpha}\; , \; K\Bigr) \;
{\boldsymbol \delta}\Bigl(\sum_{\alpha=1}^{M} s_{\alpha}\; , \; L\Bigr)
\;
\bigl( ... \bigr)
\end{equation}
which allows to lift the summations over $K$, $L$, and $\{n_{\alpha}\}$
in eq.(\ref{32}).
 In terms of the parameters $\{m_{\alpha}\}$ and $\{s_{\alpha}\}$
the product factors in eq.(\ref{36}) are expressed as follows:
\begin{eqnarray}
\label{42}
\prod_{a=1}^{K} q^{(-)}_{{\cal P}_{a}^{(K)}}
&=&
\prod_{\alpha=1}^{M} \prod_{r=1}^{m_{\alpha}}
{q^{\alpha}_{r}}^{(-)}
\\
\nonumber
\\
\label{43}
\prod_{a=1}^{L} q^{(+)}_{{\cal P}_{a}^{(L)}}
&=&
\prod_{\alpha=1}^{M} \prod_{r=1}^{s_{\alpha}}
{{q^{*}}^{\alpha}_{r}}^{(+)}
\end{eqnarray}
\begin{eqnarray}
\label{44}
\prod_{a<b}^{K}
\Biggl[
\frac{
q^{(-)}_{{\cal P}_a^{(K)}} + q^{(-)}_{{\cal P}_b^{(K)}}  + i \kappa }{
q^{(-)}_{{\cal P}_a^{(K)}} + q^{(-)}_{{\cal P}_b^{(K)}}}
\Biggr]
&=&
\prod_{\alpha=1}^{M} \prod_{1\leq r< r'}^{m_{\alpha}}
\Biggl[
\frac{
{q^{\alpha}_{r}}^{(-)}+{q^{\alpha}_{r'}}^{(-)}+i\kappa}{
{q^{\alpha}_{r}}^{(-)}+{q^{\alpha}_{r'}}^{(-)}}
\Biggr]
\times
\prod_{1\leq\alpha<\beta}^{M} \prod_{r=1}^{m_{\alpha}}\prod_{r'=1}^{m_{\beta}}
\Biggl[
\frac{
{q^{\alpha}_{r}}^{(-)}+{q^{\beta}_{r'}}^{(-)}+i\kappa}{
{q^{\alpha}_{r}}^{(-)}+{q^{\beta}_{r'}}^{(-)}}
\Biggr]
\\
\nonumber
\\
\nonumber
\\
\label{45}
\prod_{c<d}^{L}
\Biggl[
\frac{
q^{(+)}_{{\cal P}_c^{(L)}} + q^{(+)}_{{\cal P}_d^{(L)}}-i \kappa }{
q^{(+)}_{{\cal P}_c^{(L)}} + q^{(+)}_{{\cal P}_d^{(L)}}}\Biggr]
&=&
\prod_{\alpha=1}^{M} \prod_{1\leq r< r'}^{s_{\alpha}}
\Biggl[
\frac{
{{q^{*}}^{\alpha}_{r}}^{(+)}+{{q^{*}}^{\alpha}_{r'}}^{(+)}-i\kappa}{
{{q^{*}}^{\alpha}_{r}}^{(+)}+{{q^{*}}^{\alpha}_{r'}}^{(+)}}
\Biggr]
\times
\prod_{1\leq\alpha<\beta}^{M} \prod_{r=1}^{s_{\alpha}}\prod_{r'=1}^{s_{\beta}}
\Biggl[
\frac{
{{q^{*}}^{\alpha}_{r}}^{(+)}+{{q^{*}}^{\beta}_{r'}}^{(+)}-i\kappa}{
{{q^{*}}^{\alpha}_{r}}^{(+)}+{{q^{*}}^{\beta}_{r'}}^{(+)}}
\Biggr]
\\
\nonumber
\\
\nonumber
\\
\nonumber
\prod_{a=1}^{K} \prod_{c=1}^{L}
\Biggl[
\frac{
q_{{\cal P}_a^{(K)}} - q_{{\cal P}_c^{(L)}} - i\kappa}{
q_{{\cal P}_a^{(K)}} - q_{{\cal P}_c^{(L)}}}
\Biggr]
&=&
\prod_{1\leq\alpha<\beta}^{M}
\Biggl\{
\prod_{r=1}^{m_{\alpha}}\prod_{r'=1}^{s_{\beta}}
\Biggl[
\frac{
q^{\alpha}_{r} - {q^{*}}^{\beta}_{r'} - i\kappa}{
q^{\alpha}_{r} + {q^{*}}^{\beta}_{r'}  }
\Biggr]
\times
\prod_{r=1}^{s_{\alpha}}\prod_{r'=1}^{m_{\beta}}
\Biggl[
\frac{
{q^{*}}^{\alpha}_{r} - q^{\beta}_{r'} - i\kappa}{
{q^{*}}^{\alpha}_{r} - q^{\beta}_{r'}  }
\Biggr]
\Biggr\}
\times
\\
\nonumber
\\
\nonumber
\\
\label{46}
&\times&
\prod_{\alpha=1}^{M} \prod_{r=1}^{m_{\alpha}}\prod_{r'=1}^{s_{\alpha}}
\Biggl[
\frac{
q^{\alpha}_{r} - {q^{*}}^{\alpha}_{r'} - i\kappa}{
q^{\alpha}_{r} - {q^{*}}^{\alpha}_{r'}  }
\Biggr]
\end{eqnarray}

Substituting eqs.(\ref{41})-(\ref{46}) into eq.(\ref{36}),
and then substituting the resulting expression into eq.(\ref{32})
we obtain
\begin{eqnarray}
 \nonumber
W(f) &=& \lim_{\lambda\to\infty}
\Biggl\{
1 + \sum_{M=1}^{\infty} \; \frac{(-1)^{M}}{M!} \;
\prod_{\alpha=1}^{M}
\Biggl[
\sum_{m_{\alpha}+s_{\alpha}\geq 1}^{\infty}
(-1)^{m_{\alpha}+s_{\alpha}-1}
\int_{-\infty}^{+\infty} dq_{\alpha} \; 
\frac{{\cal G}\bigl(q_{\alpha}, m_{\alpha}, s_{\alpha}\bigr)}{
2\pi \kappa (m_{\alpha}+s_{\alpha})}
\times
\\
\nonumber
\\
&\times&
\mbox{\LARGE e}^{
-\frac{t}{2\beta} (m_{\alpha}+s_{\alpha}) q_{\alpha}^{2} +
\frac{\kappa^{2}}{24 \beta} (m_{\alpha}+s_{\alpha})^{3} +
\lambda (m_{\alpha}+s_{\alpha}) f}
\Biggr]
\;
|\tilde{C}_{M}({\bf q}, {\bf m + s})|^{2}
\; \prod_{1\leq\alpha<\beta}^{M} \;
{\cal G}_{\alpha\beta} \bigl({\bf q}, {\bf m}, {\bf s}\bigr)
\Biggr\}
\label{47}
\end{eqnarray}
where
\begin{equation}
\label{47a}
|\tilde{C}_{M}({\bf q}, {\bf m + s})|^{2} \; = \;
\prod_{\alpha<\beta}^{M}
\frac{
\big|q_{\alpha}-q_{\beta}-\frac{i\kappa}{2}(m_{\alpha}+s_{\alpha}-m_{\beta}-s_{\beta})\big|^{2}}{
\big|q_{\alpha}-q_{\beta}-\frac{i\kappa}{2}(m_{\alpha}+s_{\alpha}+m_{\beta}+s_{\beta})\big|^{2}}
\end{equation}
\begin{equation}
 \label{48}
{\cal G} =
\frac{(-1)^{s_{\alpha}} (-i\kappa)^{(m_{\alpha}+s_{\alpha})}}{
\prod_{r=1}^{m_{\alpha}}{q^{\alpha}_{r}}^{(-)}
\prod_{r=1}^{s_{\alpha}}{{q^{*}}^{\alpha}_{r}}^{(+)}}
\prod_{r< r'}^{m_{\alpha}}
\Biggl[
\frac{
{q^{\alpha}_{r}}^{(-)}+{q^{\alpha}_{r'}}^{(-)}+i\kappa}{
{q^{\alpha}_{r}}^{(-)}+{q^{\alpha}_{r'}}^{(-)}}
\Biggr]
\prod_{r< r'}^{s_{\alpha}}
\Biggl[
\frac{
{{q^{*}}^{\alpha}_{r}}^{(+)}+{{q^{*}}^{\alpha}_{r'}}^{(+)}-i\kappa}{
{{q^{*}}^{\alpha}_{r}}^{(+)}+{{q^{*}}^{\alpha}_{r'}}^{(+)}}
\Biggr]
\prod_{r=1}^{m_{\alpha}}\prod_{r'=1}^{s_{\alpha}}
\Biggl[
\frac{
q^{\alpha}_{r} - {q^{*}}^{\alpha}_{r'} - i\kappa}{
q^{\alpha}_{r} - {q^{*}}^{\alpha}_{r'}  }
\Biggr]
\end{equation}
and
\begin{equation}
 \label{49}
{\cal G}_{\alpha\beta} =
\prod_{r=1}^{m_{\alpha}}\prod_{r'=1}^{m_{\beta}}
\Biggl[
\frac{
{q^{\alpha}_{r}}^{(-)}+{q^{\beta}_{r'}}^{(-)}+i\kappa}{
{q^{\alpha}_{r}}^{(-)}+{q^{\beta}_{r'}}^{(-)}}
\Biggr]
\prod_{r=1}^{s_{\alpha}}\prod_{r'=1}^{s_{\beta}}
\Biggl[
\frac{
{{q^{*}}^{\alpha}_{r}}^{(+)}+{{q^{*}}^{\beta}_{r'}}^{(+)}-i\kappa}{
{{q^{*}}^{\alpha}_{r}}^{(+)}+{{q^{*}}^{\beta}_{r'}}^{(+)}}
\Biggr]
\prod_{r=1}^{m_{\alpha}}\prod_{r'=1}^{s_{\beta}}
\Biggl[
\frac{
q^{\alpha}_{r} - {q^{*}}^{\beta}_{r'} - i\kappa}{
q^{\alpha}_{r} + {q^{*}}^{\beta}_{r'}  }
\Biggr]
\times
\prod_{r=1}^{s_{\alpha}}\prod_{r'=1}^{m_{\beta}}
\Biggl[
\frac{
{q^{*}}^{\alpha}_{r} - q^{\beta}_{r'} - i\kappa}{
{q^{*}}^{\alpha}_{r} - q^{\beta}_{r'}  }
\Biggr]
\end{equation}
The product factors in eq.(\ref{48}) can be easily expressed it terms of the
Gamma functions:
\begin{eqnarray}
\label{50}
\prod_{r=1}^{m_{\alpha}}{q^{\alpha}_{r}}^{(-)} &=&
\prod_{r=1}^{m_{\alpha}}
\Bigl[
{q_{\alpha}}^{(-)} - \frac{i\kappa}{2}(m_{\alpha}+s_{\alpha}+1) + i\kappa r
\Bigr]
\; = \;
(i\kappa)^{m_{\alpha}}
\frac{
\Gamma\Bigl(
\frac{1}{2}-\frac{s_{\alpha}-m_{\alpha}}{2} - \frac{i{q_{\alpha}}^{(-)}}{\kappa}
\Bigr)}{
\Gamma\Bigl(
\frac{1}{2}-\frac{s_{\alpha}+m_{\alpha}}{2} - \frac{i{q_{\alpha}}^{(-)}}{\kappa}
\Bigr)}
\\
\nonumber
\\
\nonumber
\\
\prod_{r=1}^{s_{\alpha}}{{q^{*}}^{\alpha}_{r}}^{(+)} &=&
\prod_{r=1}^{s_{\alpha}}
\Bigl[
{q_{\alpha}}^{(+)} + \frac{i\kappa}{2}(m_{\alpha}+s_{\alpha}+1) - i\kappa r
\Bigr]
\; = \;
(-i\kappa)^{s_{\alpha}}
\frac{
\Gamma\Bigl(
\frac{1}{2}-\frac{m_{\alpha}-s_{\alpha}}{2} + \frac{i{q_{\alpha}}^{(+)}}{\kappa}
\Bigr)}{
\Gamma\Bigl(
\frac{1}{2}-\frac{m_{\alpha}+s_{\alpha}}{2} + \frac{i{q_{\alpha}}^{(+)}}{\kappa}
\Bigr)}
\label{51}
\end{eqnarray}
\begin{eqnarray}
\label{52}
\prod_{r< r'}^{m_{\alpha}}
\Biggl[
\frac{
{q^{\alpha}_{r}}^{(-)}+{q^{\alpha}_{r'}}^{(-)}+i\kappa}{
{q^{\alpha}_{r}}^{(-)}+{q^{\alpha}_{r'}}^{(-)}}
\Biggr]
&=&
2^{-(m_{\alpha}-1)}
\frac{
\Gamma\Bigl(
m_{\alpha}-s_{\alpha} - \frac{2i{q_{\alpha}}^{(-)}}{\kappa}
\Bigr)
\Gamma\Bigl(
1-\frac{m_{\alpha}+s_{\alpha}}{2} - \frac{i{q_{\alpha}}^{(-)}}{\kappa}
\Bigr)}{
\Gamma\Bigl(
\frac{m_{\alpha}-s_{\alpha}}{2} - \frac{i{q_{\alpha}}^{(-)}}{\kappa}
\Bigr)
\Gamma\Bigl(
1 - s_{\alpha} - \frac{2i{q_{\alpha}}^{(-)}}{\kappa}
\Bigr)}
\\
\nonumber
\\
\nonumber
\\
\label{53}
\prod_{r< r'}^{s_{\alpha}}
\Biggl[
\frac{
{{q^{*}}^{\alpha}_{r}}^{(+)}+{{q^{*}}^{\alpha}_{r'}}^{(+)}-i\kappa}{
{{q^{*}}^{\alpha}_{r}}^{(+)}+{{q^{*}}^{\alpha}_{r'}}^{(+)}}
\Biggr]
&=&
2^{-(s_{\alpha}-1)}
\frac{
\Gamma\Bigl(
s_{\alpha}-m_{\alpha} + \frac{2i{q_{\alpha}}^{(+)}}{\kappa}
\Bigr)
\Gamma\Bigl(
1-\frac{m_{\alpha}+s_{\alpha}}{2} + \frac{i{q_{\alpha}}^{(+)}}{\kappa}
\Bigr)}{
\Gamma\Bigl(
\frac{s_{\alpha}-m_{\alpha}}{2} + \frac{i{q_{\alpha}}^{(+)}}{\kappa}
\Bigr)
\Gamma\Bigl(
1 - m_{\alpha} + \frac{2i{q_{\alpha}}^{(+)}}{\kappa}
\Bigr)}
\\
\nonumber
\\
\nonumber
\\
\label{54}
\prod_{r=1}^{m_{\alpha}}\prod_{r'=1}^{s_{\alpha}}
\Biggl[
\frac{
q^{\alpha}_{r} - {q^{*}}^{\alpha}_{r'} - i\kappa}{
q^{\alpha}_{r} - {q^{*}}^{\alpha}_{r'}  }
\Biggr]
&=&
\frac{
\Gamma\bigl(1 + m_{\alpha} + s_{\alpha}\bigr)}{
\Gamma\bigl(1 + m_{\alpha}\bigr) \Gamma\bigl(1 + s_{\alpha}\bigr)}
\end{eqnarray}
Substituting the above expressions into eq.(\ref{48}) and using the
standard relations for the Gamma functions,
\begin{eqnarray}
\label{55}
\Gamma(z) \,\Gamma(1-z) &=& \frac{\pi}{\sin(\pi z)}
\\
\nonumber
\\
\label{56}
\Gamma(1+z) &=& z \, \Gamma(z)
\\
\nonumber
\\
\label{57}
\Gamma\Bigl(\frac{1}{2} + z\Bigr) &=&
\frac{\sqrt{\pi} \, \Gamma\bigl(1 + 2z\bigr)}{
2^{2z} \, \Gamma\bigl(1 + z\bigr)}
\end{eqnarray}
for the factor ${\cal G}$, eq.(\ref{48}), we get
\begin{equation}
\label{58}
{\cal G}\bigl(q_{\alpha}, m_{\alpha}, s_{\alpha}\bigr) \; = \;
\frac{
\Gamma\Bigl(
s_{\alpha} + \frac{2i}{\kappa} {q_{\alpha}}^{(-)}
\Bigr) \,
\Gamma\Bigl(
m_{\alpha} - \frac{2i}{\kappa} {q_{\alpha}}^{(+)}
\Bigr) \,
\Gamma\bigl(1 + m_{\alpha} + s_{\alpha}\bigr)}{
2^{(m_{\alpha} + s_{\alpha})}
\Gamma\Bigl(
m_{\alpha} + s_{\alpha} + \frac{2i}{\kappa} {q_{\alpha}}^{(-)}
\Bigr) \,
\Gamma\Bigl(
m_{\alpha} + s_{\alpha} - \frac{2i}{\kappa} {q_{\alpha}}^{(+)}
\Bigr) \,
\Gamma\bigl(1 + m_{\alpha}\bigr) \Gamma\bigl(1 + s_{\alpha}\bigr)}
\end{equation}
Similar calculations for the factor ${\cal G}_{\alpha\beta}$
yield the following expression
\begin{eqnarray}
\nonumber
{\cal G}_{\alpha\beta} \bigl({\bf q}, {\bf m}, {\bf s}\bigr) &=&
\frac{
\Gamma
\Bigl[
1 + \frac{m_{\alpha} + m_{\beta} - s_{\alpha} - s_{\beta}}{2}
-\frac{i}{\kappa}\bigl({q_{\alpha}}^{(-)} + {q_{\beta}}^{(-)}\bigr)
\Bigr] \,
\Gamma
\Bigl[
1 - \frac{m_{\alpha} + m_{\beta} + s_{\alpha} + s_{\beta}}{2}
-\frac{i}{\kappa}\bigl({q_{\alpha}}^{(-)} + {q_{\beta}}^{(-)}\bigr)
\Bigr]}{
\Gamma
\Bigl[
1 - \frac{m_{\alpha} - m_{\beta} + s_{\alpha} + s_{\beta}}{2}
-\frac{i}{\kappa}\bigl({q_{\alpha}}^{(-)} + {q_{\beta}}^{(-)}\bigr)
\Bigr] \,
\Gamma
\Bigl[
1 + \frac{m_{\alpha} - m_{\beta} - s_{\alpha} - s_{\beta}}{2}
-\frac{i}{\kappa}\bigl({q_{\alpha}}^{(-)} + {q_{\beta}}^{(-)}\bigr)
\Bigr]}
\times
\\
\nonumber
\\
\nonumber
\\
\nonumber
&\times&
\frac{
\Gamma
\Bigl[
1 - \frac{m_{\alpha} + m_{\beta} - s_{\alpha} - s_{\beta}}{2}
+\frac{i}{\kappa}\bigl({q_{\alpha}}^{(+)} + {q_{\beta}}^{(+)}\bigr)
\Bigr] \,
\Gamma
\Bigl[
1 - \frac{m_{\alpha} + m_{\beta} + s_{\alpha} + s_{\beta}}{2}
+\frac{i}{\kappa}\bigl({q_{\alpha}}^{(+)} + {q_{\beta}}^{(+)}\bigr)
\Bigr]}{
\Gamma
\Bigl[
1 - \frac{m_{\alpha} + m_{\beta} + s_{\alpha} - s_{\beta}}{2}
+\frac{i}{\kappa}\bigl({q_{\alpha}}^{(+)} + {q_{\beta}}^{(+)}\bigr)
\Bigr] \,
\Gamma
\Bigl[
1 - \frac{m_{\alpha} + m_{\beta} - s_{\alpha} + s_{\beta}}{2}
+\frac{i}{\kappa}\bigl({q_{\alpha}}^{(+)} + {q_{\beta}}^{(+)}\bigr)
\Bigr]}
\times
\\
\nonumber
\\
\nonumber
\\
\nonumber
&\times&
\frac{
\Gamma
\Bigl[
1 + \frac{m_{\alpha} + m_{\beta} + s_{\alpha} + s_{\beta}}{2}
+\frac{i}{\kappa}\bigl(q_{\alpha} - q_{\beta}\bigr)
\Bigr] \,
\Gamma
\Bigl[
1 + \frac{-m_{\alpha} + m_{\beta} + s_{\alpha} - s_{\beta}}{2}
+\frac{i}{\kappa}\bigl(q_{\alpha} - q_{\beta}\bigr)
\Bigr]}{
\Gamma
\Bigl[
1 + \frac{-m_{\alpha} + m_{\beta} + s_{\alpha} + s_{\beta}}{2}
+\frac{i}{\kappa}\bigl(q_{\alpha} - q_{\beta}\bigr)
\Bigr] \,
\Gamma
\Bigl[
1 + \frac{m_{\alpha} + m_{\beta} + s_{\alpha} - s_{\beta}}{2}
+\frac{i}{\kappa}\bigl(q_{\alpha} - q_{\beta}\bigr)
\Bigr]}
\times
\\
\nonumber
\\
\nonumber
\\
\label{59}
&\times&
\frac{
\Gamma
\Bigl[
1 + \frac{m_{\alpha} + m_{\beta} + s_{\alpha} + s_{\beta}}{2}
-\frac{i}{\kappa}\bigl(q_{\alpha} - q_{\beta}\bigr)
\Bigr] \,
\Gamma
\Bigl[
1 + \frac{m_{\alpha} - m_{\beta} - s_{\alpha} + s_{\beta}}{2}
-\frac{i}{\kappa}\bigl(q_{\alpha} - q_{\beta}\bigr)
\Bigr]}{
\Gamma
\Bigl[
1 + \frac{m_{\alpha} + m_{\beta} - s_{\alpha} + s_{\beta}}{2}
-\frac{i}{\kappa}\bigl(q_{\alpha} - q_{\beta}\bigr)
\Bigr] \,
\Gamma
\Bigl[
1 + \frac{m_{\alpha} - m_{\beta} + s_{\alpha} + s_{\beta}}{2}
-\frac{i}{\kappa}\bigl(q_{\alpha} - q_{\beta}\bigr)
\Bigr]}
\end{eqnarray}

\vspace{10mm}

Redefining
\begin{equation}
\label{60}
q_{\alpha} \; = \; \frac{\kappa}{2\lambda} \, p_{\alpha}
\end{equation}
with
\begin{equation}
\label{61}
\lambda \; = \; \frac{1}{2} \,
\Bigl(\frac{\kappa^{2} t}{\beta}\Bigr)^{1/3} \; = \;
\frac{1}{2} \, \bigl(\beta^{5} u^{2} t\bigr)^{1/3}
\end{equation}
the normalization factor $|\tilde{C}_{M}({\bf q}, {\bf m + s})|^{2}$, eq.(\ref{47a}),
can be represented as follows:
\begin{eqnarray}
\nonumber
|\tilde{C}_{M}({\bf q}, {\bf m + s})|^{2} &=&
\prod_{\alpha<\beta}^{M}
\frac{
\big|
\lambda\bigl(m_{\alpha}+s_{\alpha}\bigr) - \lambda\bigl(m_{\beta}+s_{\beta}\bigr) -
i p_{\alpha} + ip_{\beta}
\big|^{2}}{
\big|
\lambda\bigl(m_{\alpha}+s_{\alpha}\bigr) + \lambda\bigl(m_{\beta}+s_{\beta}\bigr) -
i p_{\alpha} + ip_{\beta}
\big|^{2} }
\; = \;
\\
\nonumber
\\
\label{62}
&=&
\prod_{\alpha=1}^{M}
\bigl[2\lambda \bigl(m_{\alpha}+s_{\alpha}\bigr)\bigr]
\times
\det
\Biggl[
\frac{1}{
\lambda\bigl(m_{\alpha}+s_{\alpha}\bigr) - ip_{\alpha} +
\lambda\bigl(m_{\beta}+s_{\beta}\bigr) + ip_{\beta}}
\Biggr]_{\alpha,\beta=1,...,M}
\end{eqnarray}
where we have used the Cauchy double alternant identity
\begin{equation}
 \label{63}
\frac{\prod_{\alpha<\beta}^{M} (a_{\alpha} - a_{\beta})(b_{\alpha} - b_{\beta})}{
     \prod_{\alpha,\beta=1}^{M} (a_{\alpha} - b_{\beta})} \; = \;
(-1)^{M(M-1)/2} \det\Bigl[\frac{1}{a_{\alpha}-b_{\beta}}\Bigr]_{\alpha,\beta=1,...M}
\end{equation}
with $a_{\alpha} = p_{\alpha} - i \lambda \bigl(m_{\alpha}+s_{\alpha}\bigr)$ and
$b_{\alpha} = p_{\alpha} + i \lambda \bigl(m_{\beta}+s_{\beta}\bigr)$.

After rescaling, eq.(\ref{60}), for the exponential factor in eq.(\ref{47}) we find
\begin{equation}
\label{64}
-\frac{t}{2\beta} (m_{\alpha}+s_{\alpha}) q_{\alpha}^{2}
+\frac{\kappa^{2}}{24 \beta} (m_{\alpha}+s_{\alpha})^{3}
+\lambda (m_{\alpha}+s_{\alpha}) f
\; = \;
-\lambda (m_{\alpha}+s_{\alpha}) p_{\alpha}^{2}
+\frac{1}{3} \lambda^{3} (m_{\alpha}+s_{\alpha})^{3}
+\lambda (m_{\alpha}+s_{\alpha}) f
\end{equation}
The cubic exponential term can be linearized using the Airy function relation
\begin{equation}
   \label{65}
\exp\Bigl[ \frac{1}{3} \lambda^{3} (m_{\alpha}+s_{\alpha})^{3} \Bigr] \; = \;
\int_{-\infty}^{+\infty} dy_{\alpha} \; \Ai(y_{\alpha}) \;
\exp\Bigl[\lambda (m_{\alpha}+s_{\alpha}) \, y_{\alpha} \Bigr]
\end{equation}
Substituting eqs.(\ref{65}),(\ref{64}) and (\ref{62}) into eq.(\ref{47}), and redefining
$y_{\alpha} \; \to \; y_{\alpha} + p_{\alpha}^{2} - f$, we get
\begin{eqnarray}
 \nonumber
W(f) \; = \; \lim_{\lambda\to\infty}
\Biggl\{
&1& + \sum_{M=1}^{\infty} \; \frac{(-1)^{M}}{M!} \;
\prod_{\alpha=1}^{M}
\Biggl[
\int\int_{-\infty}^{+\infty} \frac{dy_{\alpha} dp_{\alpha}}{2\pi}
\Ai\bigl(y_{\alpha} + p_{\alpha}^{2} - f \bigr)
\times
\\
\nonumber
\\
\nonumber
&\times&
\sum_{m_{\alpha}+s_{\alpha}\geq 1}^{\infty} (-1)^{m_{\alpha}+s_{\alpha}-1}
\exp\{\lambda (m_{\alpha}+s_{\alpha})y_{\alpha}\} \;
{\cal G} \Bigl(\frac{p_{\alpha}}{\lambda}, \; m_{\alpha}, \; s_{\alpha}\Bigr) \;
2^{m_{\alpha}+s_{\alpha}}
\Biggr]
\times
\\
\nonumber
\\
&\times&
\det \hat{K}\bigl[(\lambda m_{\alpha},\, \lambda s_{\alpha}, \, p_{\alpha});
(\lambda m_{\beta}, \, \lambda s_{\beta}, \, p_{\beta})\bigr]_{\alpha,\beta=1,...,M}
\times \prod_{1\leq\alpha<\beta}^{M} \;
{\cal G}_{\alpha\beta}\Bigl(\frac{{\bf p}}{\lambda}, \; {\bf m}, \; {\bf s}\Bigr)
\Biggr\}
\label{66}
\end{eqnarray}
where
\begin{equation}
\label{67}
\hat{K}\bigl[(\lambda m, \, \lambda s, \, p); (\lambda m', \, \lambda s', \, p')\bigr]
\; = \;
\frac{1}{
\lambda m + \lambda s - ip +
\lambda m' + \lambda s' + ip'}
\end{equation}

The crucial point of the further calculations is the procedure of
taking the thermodynamic limit $\lambda \to \infty$. First of all one can easily note that
according to eqs.(\ref{60}) and (\ref{66}), it is the parameters
$p_{\alpha} \sim \lambda q_{\alpha}$ which remain finite in the
limit $\lambda \to \infty$. In other words, all the parameters $q_{\alpha}$
which are not multiplied by $\lambda$ (e.g. in the expressions for
${\cal G}_{\alpha}$ and ${\cal G}_{\alpha\beta}$, eqs(\ref{58}) and (\ref{59}))
have to be taken to zero in this limit. Simultaneously, the summations over $\{m_{\alpha}\}$
and $\{s_{\alpha}\}$ have to be performed. The general algorithm of such
summation is in the following. Let us consider the example of the sum of a general type:
\begin{equation}
\label{68}
R({\bf y}, {\bf p}) \; = \; \lim_{\lambda\to\infty} \prod_{\alpha=1}^{M}
\Biggl[
\sum_{n_{\alpha}=1}^{\infty} \; (-1)^{n_{\alpha} - 1}
\exp\{\lambda n_{\alpha} y_{\alpha}\}
\Biggr]\;
\Phi\Bigl[
{\bf p},\; \frac{{\bf p}}{\lambda}, \; \lambda {\bf n}; \; {\bf n}
\Bigr]
\end{equation}
where $\Phi$
is a function which depend both of $\lambda n_{\alpha}$'s
and $n_{\alpha}$'s (which are not multiplied by $\lambda$).
The summations in the above example can be represented in terms
of the integrals in the complex plane:
\begin{equation}
\label{69}
R({\bf y}, {\bf p}) \; = \; \lim_{\lambda\to\infty} \prod_{\alpha=1}^{M}
\Biggl[
\frac{1}{2i} \int_{{\cal C}} \frac{dz_{\alpha}}{\sin(\pi z_{\alpha})}
\exp\{\lambda z_{\alpha} y_{\alpha}\}
\Biggr]\;
\Phi\Bigl[
{\bf p},\; \frac{{\bf p}}{\lambda}, \; \lambda {\bf z}; \; {\bf z}
\Bigr]
\end{equation}
where the integration goes over the contour ${\cal C}$ shown in Fig.1(a).
Shifting the contour to the position ${\cal C}'$ shown in Fig.1(b) (assuming that there is no contribution from $\infty$), and redefining $z \to z/\lambda$, in the
limit $\lambda \to \infty$ we get:
\begin{equation}
\label{70}
R({\bf y}, {\bf p}) \; = \;  \prod_{\alpha=1}^{M}
\Biggl[
\frac{1}{2\pi i} \int_{{\cal C}'} \frac{dz_{\alpha}}{z_{\alpha}}
\exp\{z_{\alpha} y_{\alpha}\}
\Biggr]\;
\lim_{\lambda\to\infty}
\Phi\Bigl[
{\bf p},\; \frac{{\bf p}}{\lambda}, \; {\bf z}; \; \frac{{\bf z}}{\lambda}
\Bigr]
\end{equation}
where the parameters $y_{\alpha}$, $p_{\alpha}$ and $z_{\alpha}$
remain finite in the limit $\lambda \to \infty$.

\begin{figure}[h]
\begin{center}
   \includegraphics[width=12.0cm]{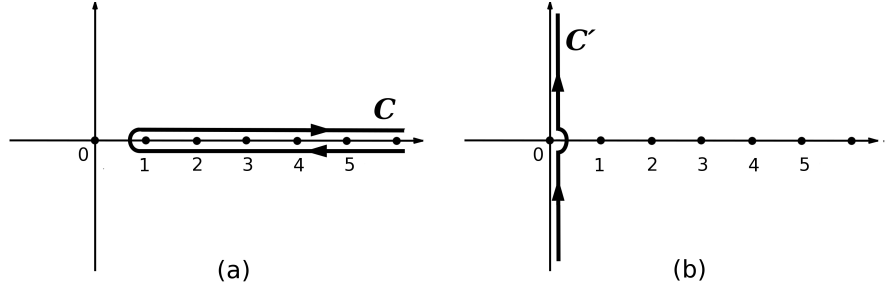}
\caption[]{The contours of integration in the complex plane used for
           summing the series:
           (a) the original contour ${\cal C}$;
           (b) the deformed contour ${\cal C}'$; }
\end{center}
\label{figure1}
\end{figure}

Let us consider now the summations in eq.(\ref{66}).
Here the double sum can be represented as follows:
\begin{eqnarray}
\nonumber
\sum_{m_{\alpha}+s_{\alpha}\geq 1}^{\infty} (-1)^{m_{\alpha}+s_{\alpha}-1}
\; f\bigl({\bf m}; {\bf s}\bigr)
&=&
\sum_{m_{\alpha}=1}^{\infty} (-1)^{m_{\alpha}-1}
\; f\bigl({\bf m}; {\bf s}\bigr)\big|_{s_{\alpha}=0} \;+ \;
\sum_{s_{\alpha}=1}^{\infty} (-1)^{s_{\alpha}-1}
\; f\bigl({\bf m}; {\bf s}\bigr)\big|_{m_{\alpha}=0} \;- \;
\\
\nonumber
\\
&-&
\sum_{m_{\alpha}=1}^{\infty} (-1)^{m_{\alpha}-1}
\sum_{s_{\alpha}=1}^{\infty} (-1)^{s_{\alpha}-1}
f\bigl({\bf m}; {\bf s}\bigr)
\label{71}
\end{eqnarray}
Thus, according to the above summation algorithm, we get
\begin{equation}
\label{72}
\lim_{\lambda\to\infty}
\sum_{m_{\alpha}+s_{\alpha}\geq 1}^{\infty} (-1)^{m_{\alpha}+s_{\alpha}-1}
\; f\bigl({\bf m}; {\bf s}\bigr)
 =
\frac{1}{(2\pi i)^{2}}
\int\int_{{\cal C}'}
\frac{d{z_{1}}_{\alpha}}{{z_{1}}_{\alpha}}
\frac{d{z_{2}}_{\alpha}}{{z_{2}}_{\alpha}}
\Bigl[
(2\pi i) \delta({z_{2}}_{\alpha}) + (2\pi i) \delta({z_{1}}_{\alpha}) - 1
\Bigr]
\lim_{\lambda\to\infty}
f\Bigl(\frac{{\bf z}_{1}}{\lambda}; \frac{{\bf z}_{2}}{\lambda}\Bigr)
\end{equation}
where the rescaled integration parameters
${z_{1}}_{\alpha}$ and
${z_{2}}_{\alpha}$
remain finite in the limit $\lambda \to \infty$.
Finally, taking into account the Gamma function properties,
$\Gamma(z)|_{|z|\to 0} = 1/z$ and
$\Gamma(1+z)|_{|z|\to 0} = 1$, for the factors ${\cal G}$ and
${\cal G}_{\alpha\beta}$, eqs.(\ref{58})-(\ref{59}), we easily find
\begin{eqnarray}
\nonumber
\lim_{\lambda\to\infty}
{\cal G}\bigl(q_{\alpha}, m_{\alpha}, s_{\alpha}\bigr) \;
2^{(m_{\alpha}+s_{\alpha})}
&=&
\lim_{\lambda\to\infty}
{\cal G}\Bigl(
\frac{p_{\alpha}}{\lambda}, \;
\frac{{z_{1}}_{\alpha}}{\lambda} , \;
\frac{{z_{2}}_{\alpha}}{\lambda}
\Bigr) 2^{({z_{1}}_{\alpha}/\lambda + {z_{2}}_{\alpha}/\lambda)}
\\
\nonumber
\\
&=&
\frac{
\bigl({z_{1}}_{\alpha}+{z_{2}}_{\alpha} + i p_{\alpha}^{(-)}\bigr)
\bigl({z_{1}}_{\alpha}+{z_{2}}_{\alpha} - i p_{\alpha}^{(+)}\bigr)}{
\bigl({z_{2}}_{\alpha} + i p_{\alpha}^{(-)}\bigr)
\bigl({z_{1}}_{\alpha} - i p_{\alpha}^{(+)}\bigr)}
\; \equiv \;
{\cal G}_{*}\bigl(p_{\alpha}, {z_{1}}_{\alpha}, {z_{2}}_{\alpha}\bigr)
\label{73}
\end{eqnarray}
and
\begin{equation}
\label{74}
\lim_{\lambda\to\infty}
{\cal G}_{\alpha\beta}\Bigl(\frac{{\bf p}}{\lambda}, \; {\bf m}, \; {\bf s}\Bigr)
\; = \;
\lim_{\lambda\to\infty}
{\cal G}_{\alpha\beta}
\Bigl(
\frac{{\bf p}}{\lambda}, \; \frac{{\bf z}_{1}}{\lambda}, \;
\frac{{\bf z}_{2}}{\lambda}\Bigr)
\; = \;
1
\end{equation}
where
\begin{equation}
\label{75}
p_{\alpha}^{(\pm)} \; = \; p_{\alpha} \; \pm \; i \epsilon
\end{equation}
Thus, in the limit $\lambda \to \infty$ the expression for the probability distribution function, eq.(\ref{66}),
takes the form of the Fredholm determinant
\begin{eqnarray}
 \nonumber
W(f) &=&
1 + \sum_{M=1}^{\infty} \; \frac{(-1)^{M}}{M!} \;
\prod_{\alpha=1}^{M}
\Biggl[
\int\int_{-\infty}^{+\infty} \frac{dy_{\alpha} dp_{\alpha}}{2\pi}
\Ai\bigl(y_{\alpha} + p_{\alpha}^{2} - f \bigr)
\\
\nonumber
\\
\nonumber
&\times&
\frac{1}{(2\pi i)^{2}}
\int\int_{{\cal C}'}
\frac{d{z_{1}}_{\alpha}}{{z_{1}}_{\alpha}}
\frac{d{z_{2}}_{\alpha}}{{z_{2}}_{\alpha}}
\Bigl[
(2\pi i) \delta({z_{2}}_{\alpha}) + (2\pi i) \delta({z_{1}}_{\alpha}) - 1
\Bigr]
\Bigl(
1 + \frac{{z_{1}}_{\alpha}}{{z_{2}}_{\alpha} + i p_{\alpha}^{(-)}}
\Bigr)
\Bigl(
1 + \frac{{z_{2}}_{\alpha}}{{z_{1}}_{\alpha} - i p_{\alpha}^{(+)}}
\Bigr) \;
\mbox{\LARGE e}^{({z_{1}}_{\alpha} + {z_{2}}_{\alpha}) y_{\alpha}}
\Biggr]
\\
\nonumber
\\
\nonumber
&\times&
\det \hat{K}\bigl[({z_{1}}_{\alpha},{z_{2}}_{\alpha},p_{\alpha});
({z_{1}}_{\beta},{z_{2}}_{\beta},p_{\beta})\bigr]_{\alpha,\beta=1,...,M}
\\
\nonumber
\\
&=&
\det\bigl[1 \, - \, \hat{K} \bigr]
\label{76}
\end{eqnarray}
with the kernel
\begin{equation}
\label{77}
\hat{K}\bigl[({z_{1}}, \, {z_{2}}, \, p); ({z_{1}}', \, {z_{2}}', \, p')\bigr]
\; = \;
\frac{1}{
{z_{1}} + {z_{2}} - ip +
{z_{1}}' + {z_{2}}' + ip'}
\end{equation}
In the exponential representation of this determinant we get
\begin{equation}
 \label{78}
W(f) \; = \;
\exp\Bigl[-\sum_{M=1}^{\infty} \frac{1}{M} \; \mbox{Tr} \, \hat{K}^{M} \Bigr]
\end{equation}
where
\begin{eqnarray}
 \nonumber
\mbox{Tr} \, \hat{K}^{M} &=&
\prod_{\alpha=1}^{M}
\Biggl[
\int\int_{-\infty}^{+\infty} \frac{dy_{\alpha} dp_{\alpha}}{2\pi}
\Ai\bigl(y_{\alpha} + p_{\alpha}^{2} - f \bigr)
\times
\\
\nonumber
\\
\nonumber
&\times&
\frac{1}{(2\pi i)^{2}}
\int\int_{{\cal C}'}
\frac{d{z_{1}}_{\alpha}}{{z_{1}}_{\alpha}}
\frac{d{z_{2}}_{\alpha}}{{z_{2}}_{\alpha}}
\Bigl[
(2\pi i) \delta({z_{2}}_{\alpha}) + (2\pi i) \delta({z_{1}}_{\alpha}) - 1
\Bigr]
\Bigl(
1 + \frac{{z_{1}}_{\alpha}}{{z_{2}}_{\alpha} + i p_{\alpha}^{(-)}}
\Bigr)
\Bigl(
1 + \frac{{z_{2}}_{\alpha}}{{z_{1}}_{\alpha} - i p_{\alpha}^{(+)}}
\Bigr) \;
\mbox{\LARGE e}^{ ({z_{1}}_{\alpha} + {z_{2}}_{\alpha}) y_{\alpha}}
\Biggr]
\times
\\
\nonumber
\\
&\times&
\prod_{\alpha=1}^{M}
\Biggl[
\frac{1}{
{z_{1}}_{\alpha} + {z_{2}}_{\alpha} - i p_{\alpha} +
{z_{1}}_{\alpha +1}  + {z_{2}}_{\alpha +1}  + i p_{\alpha +1}}
\Biggr]
\label{79}
\end{eqnarray}
Here, by definition, it is assumed that ${z_{i_{M +1}}} \equiv {z_{i_{1}}}$ ($i=1,2$)
and $p_{M +1} \equiv p_{1}$.
Substituting
\begin{equation}
\label{80}
\frac{1}{
{z_{1}}_{\alpha} + {z_{2}}_{\alpha} - i p_{\alpha} +
{z_{1}}_{\alpha +1}  + {z_{2}}_{\alpha +1}  + i p_{\alpha +1}}
\; = \;
\int_{0}^{\infty} d\omega_{\alpha}
\exp\Bigl[-\bigl(
{z_{1}}_{\alpha} + {z_{2}}_{\alpha} - i p_{\alpha} +
{z_{1}}_{\alpha +1}  + {z_{2}}_{\alpha +1}  + i p_{\alpha +1}
\bigr) \, \omega_{\alpha}
\Bigr]
\end{equation}
into eq.(\ref{79}), we obtain
\begin{equation}
 \label{81}
\mbox{Tr} \, \hat{K}^{M} \; = \;
\int_{0}^{\infty} d\omega_{1} \, ... \, d\omega_{M} \,
\prod_{\alpha=1}^{M}
\Biggl[
\int\int_{-\infty}^{+\infty} \frac{dy_{\alpha} dp_{\alpha}}{2\pi}
\Ai\bigl(y_{\alpha} + p_{\alpha}^{2} + \omega_{\alpha} + \omega_{\alpha -1} - f \bigr)
\exp\{i p_{\alpha} \bigl(\omega_{\alpha} - \omega_{\alpha -1}\bigr)\}
\;
S\bigl(p_{\alpha}, \, y_{\alpha} \bigr)
\Biggr]
\end{equation}
where, by definition, $\omega_{0} \equiv \omega_{M}$, and
\begin{equation}
\label{82}
S\bigl(p, \, y \bigr) \; = \;
\frac{1}{(2\pi i)^{2}}
\int\int_{{\cal C}'}
\frac{d z_{1}}{z_{1}} \,
\frac{d z_{2}}{z_{2}}
\Bigl[
(2\pi i) \delta(z_{2}) + (2\pi i) \delta(z_{1}) - 1
\Bigr]
\Bigl(
1 + \frac{z_{1}}{z_{2} + i p^{(-)}}
\Bigr)
\Bigl(
1 + \frac{z_{2}}{z_{1} - i p^{(+)}}
\Bigr) \;
\mbox{\LARGE e}^{(z_{1} + z_{2}) y_{\alpha}}
\end{equation}
Simple calculations yield:
\begin{eqnarray}
\nonumber
S\bigl(p, \, y \bigr) &=&
\frac{1}{2\pi i}
\int_{{\cal C}'}
\frac{dz_{1}}{z_{1}} \,
\Bigl(
1 + \frac{z_{1}}{ i (p-i\epsilon)}
\Bigr) \,
\exp\{ z_{1}  y\}
\; + \;
\frac{1}{2\pi i}
\int_{{\cal C}'}
\frac{d z_{2}}{z_{2}} \,
\Bigl(
1 - \frac{z_{2}}{ i (p+i\epsilon)}
\Bigr) \,
\exp\{ z_{2}  y\} \; -
\\
\nonumber
\\
\nonumber
&-&
\frac{1}{(2\pi i)^{2}}
\int\int_{{\cal C}'}
\frac{dz_{1}}{z_{1}} \,
\frac{dz_{2}}{z_{2}}
\Bigl(
1 + \frac{z_{1}}{z_{2} + i (p-i\epsilon)}
\Bigr)
\Bigl(
1 + \frac{z_{2}}{z_{1} - i (p+i\epsilon)}
\Bigr)
\exp\{ (z_{1} + z_{2}) y\}
\\
\nonumber
\\
\nonumber
\\
\label{83}
&=&
\Bigl[\frac{1}{ i (p-i\epsilon)} - \frac{1}{ i (p+i\epsilon)}\Bigr] \;
\delta(y)
\end{eqnarray}
Taking the limit $\epsilon \to 0$ we find:
\begin{equation}
\label{84}
S\bigl(p, \, y \bigr) \; = \; \delta(y) \delta(p)
\end{equation}
Substituting this result into eq.(\ref{81}) we obtain
\begin{equation}
 \label{85}
\mbox{Tr} \, \hat{K}^{M} \; = \;
\int_{0}^{\infty} d\omega_{1} \, ... \, d\omega_{M} \,
\prod_{\alpha=1}^{M}
\Bigl[
\Ai\bigl( \omega_{\alpha} + \omega_{\alpha -1} - f \bigr)
\Bigr]
\end{equation}
In other words, the free energy distribution function of our problem
is given by the Fredholm determinant,
\begin{equation}
\label{86}
W(f) \; = \; \det\Bigl[1 \; - \; \hat{K}_{-f}\Bigr]
\end{equation}
with the kernel
\begin{equation}
\label{87}
K_{-f}(\omega, \omega') \; =\;
\Ai\bigl( \omega \; + \; \omega' \; - \; f \bigr) \; ,
\; \; \; \; \; \; \; \; \; \; \; \; \; \; \; \; (\omega , \; \omega' \; > \; 0)
\end{equation}
which is the GOE Tracy-Widom distribution \cite{TW2,Ferrari-Spohn}.

\section{Conclusions}

In this paper we have presented sufficiently simple derivation of the
GOE Tracy-Widom distribution function for the free energy fluctuations in
random directed polymers with free boundary conditions.
The main message of this somewhat technical work is not
the final result itself (which is not new anyway), but the demonstration
of the efficiency of the general method and  new
technical tricks used in the derivation.
By mapping the original problem to the $N$-particle quantum boson system
with attractive interactions the derivation is done in the framework
of the integer replica series summations and the Bethe ansatz
formalism for the quantum boson system.

The key technical tricks of presented calculations includes the following points.
First of all, to make the integration over particle coordinates of the Bethe ansatz
propagator well defined one has to introduce proper
regularization at $\pm\infty$ which requires formal
splitting the partition function into two parts:
the one in the positive particles coordinates
sector (up to $+\infty$) and another one in
the negative particles coordinates sector (down to $-\infty$),
eqs.(\ref{14})-(\ref{15}).
Next is the "magic" Bethe ansatz combinatorial identity, eq.(\ref{35}),
which allows to perform the summation over the momenta permutations and
"disentangle" sophisticated  products containing in the Bethe ansatz
propagator. One more trick is the reformulation of the summation over
permutations of the momenta between the positive and the negative
particles positions sectors in terms of the series summations, eq.(\ref{41}),
which allows to represent the probability distribution function
in terms of the problem of the series summations, eq.(\ref{47}).
Finally, the crucial point of the considered derivation is
the procedure of the series summations in the thermodynamic
limit $t \to \infty$.  In this limit, due to the integral representation
of the series, eqs.(\ref{68})-(\ref{70}), one obtains dramatic simplifications
of some factors, eqs.(\ref{73})-(\ref{74}), in the expression for
the probability distribution function,
which allows to represent it in the form of the Fredholm determinant,
eq.(\ref{76}).

Hopefully, the experience obtained in the presented calculations would help
to solve more serious long standing problems of this scope, such as
the distribution function of the directed polymer's end point fluctuations or
the statistical properties of the free energy fluctuations at different times.

\acknowledgments

An essential part of this work was done
during my visit to the Fields Institute, Toronto, in the spring of 2011
in the framework of the program "Dynamics and transport in disordered systems".
I would like to thank Jeremy Quastel and Kostya Khanin for numerous 
illuminating discussions which were crucial for the progress in these somewhat
complicated calculations.

This work was supported in part by the grant IRSES DCPA PhysBio-269139.



\end{document}